\documentclass[aps,prb,reprint,superscriptaddress,preprintnumber,showpacs]{revtex4-1}
\usepackage{graphicx}
\usepackage{color}
\usepackage{dcolumn}

\begin{document}


\title{Large magnetothermopower and Fermi surface reconstruction in Sb$_2$Te$_2$Se}
\author{Kefeng Wang}
\affiliation{Condensed Matter Physics and Materials Science Department, Brookhaven National Laboratory, Upton, New York 11973, USA}
\author{D. Graf}
\affiliation{National High Magnetic Field Laboratory, Florida State University, Tallahassee, Florida 32306-4005, USA}
\author{C. Petrovic}
\affiliation{Condensed Matter Physics and Materials Science Department, Brookhaven National Laboratory, Upton, New York 11973, USA}

\date{\today}

\begin{abstract}
We report the magnetoresistance, magnetothermopower and quantum oscillation study of Sb$_2$Te$_2$Se single crystal.  The in-plane transverse magnetoresistance
exhibits a crossover at a critical field $B^*$ from semiclassical weak-field $B^2$ dependence to the high-field unsaturated linear magnetoresistance which persists up to the
room temperature. The low-temperature Seebeck coefficient is negative in zero field contrary
to the positive Hall resistivity, indicating the multiband effect. The magnetic field induced the sign reversion of the Seebeck coefficient between 2 K and 150 K,
. The quantum oscillation of crystals
reveals the quasi-two-dimensional (quasi-2D) Fermi surface. These effects are possibly attributed to the large Fermi surface which touches Brillouin zone boundary to becomes quasi-2D and the variation in the chemical potential induced by the magnetic field.
\end{abstract}

\maketitle

\section{Introduction}
The VA-VIA compounds (such as Bi$_2$Te$_3$, Sb$_2$Te$_3$) are well known and have been extensively studied long time ago for their excellent thermoelectric
properties.\cite{thermoelectric1,thermoelectric2} Recently a new property of these materials, topological insulator (TI), became one of the mostly focused subject
of the condensed matter physics and material science.\cite{TI1,TI2} In the three dimensional (3D) TIs, the existence of the nontrivial topological invariance dictates
that the excitation gap must vanish at the boundaries, thereby inducing the robust metallic surface states in contrast to the full insulating gap in the bulk.
The spin-momentum-locked gapless surface states consisting of spin helical Dirac fermions with a quantum Berry phase could result in a variety of interesting
quantum phenomena, such as the quantum anomalous Hall effect and topological superconductor, because the back-scattering and localizaton is suppressed.\cite{TI1,TI2,TI3,TI4,TI5}
A number of materials (including HgTe quantum well, Bi$_{1-x}$Sb$_x$, Bi$_2$Se$_3$, Bi$_2$Te$_3$, and Sb$_2$Te$_3$) have been identified to be 3D TIs,
both through theoretical calculations and angular resolved photoemission spectroscopy (ARPES).\cite{TI6,TI7,TI8,TI9,TI10} Among them, the tetradymite-like compounds
such as Bi$_2$Te$_2$Se, Bi$_2$Te$_2$S, Bi$_2$Se$_2$S and Sb$_2$Te$_2$Se were predicted to host an isolated Dirac cone on their naturally cleaved surface and attracted
intensive attention.\cite{221-1,221-2,221-3,221-4}

Besides the intensive study of the surface state in TIs by ARPES, scanning tunneling microscopy (STM) and theoretical calculation, the magnetotransport behavior of bulk TIs is also important since it is directly related to the practical application. The gapless surface Dirac states will induce some quantum
transport behavior in TIs that results in very large linear magnetoresistance (MR~\cite{TI4,TI5,QT1} and quantum oscillation with Landau sublevels \cite{QT2}. In addition,
in the highly doped TIs which have bulk carriers and exhibit metallic behavior, the competition between the contribution of the bulk carriers and the surface Dirac carriers
could induce some interesting transport properties, such as the quantum Hall effect and layered transport of bulk carriers in doped Bi$_2$Te$_3$,\cite{metal1}, and the field-induced polarized transport of valleys in p-typed Sb$_2$Te$_3$.\cite{metal2} Especially in Bi$_2$Se$_3$ crystals with bulk carriers, thermoelectric/thermomagnetic studies reveal the large Zeeman splitting of the three-dimensional bulk band and the variation of the chemical potential above the quantum limit.\cite{behnia} These demonstrate the complexity of the surface/bulk states and the rich magnetotransport in TIs with different bulk carrier densities. Besides that, although the thermoelectric properties of VA-VIA compounds have been intensively studied, there are few reports about the magnetothermpower behavior. The magnetic field influence on the thermal transport in ordinary metals is usually very small. Initially the large magnetothermopower effect was observed in doped InSb which was attributed to the effects of the sample geometry on the minority carriers.\cite{InSb} In a system with large magnetoresistant effect the magnetic field has significant influence on the properties of carriers and large magnetothermopower effect could be expected. The giant magnetothermopower effect was achieved in the giant magnetoresistant multilayer/granular systems and the colossal magnetoresistant manganites, which could be of interest for magnetic field sensors or magnetic controllable thermoelectric devices.\cite{gmr,cmr}

In this paper, we report the magnetoresistance, magnetothermopower and quantum oscillation study of Sb$_2$Te$_2$Se single crystal.  The in-plane transverse magnetoresistance
exhibits a crossover at a critical field $B^*$ from semiclassical weak-field $B^2$ dependence to the high-field unsaturated linear magnetoresistance which persists up to the
room temperature. The low-temperature Seebeck coefficient is negative in zero field contrary
to the positive Hall resistivity, indicating the multiband effect. The magnetic field induced the sign reversion of the Seebeck coefficient between 2 K and 150 K,
. The quantum oscillation of crystals
reveals the quasi-2D Fermi surface. These effects are possibly attributed to the large Fermi surface which touches Brillouin zone boundary and becomes quasi-2D.

\section{Experimental}

Single crystals of Sb$_2$Te$_2$Se were grown using a high-temperature modified Bridgman method. X-ray diffraction (XRD) data were taken with Cu K$_{\alpha}$ ($\lambda=0.15418$ nm)
radiation of Rigaku Miniflex powder diffractometer. Electrical transport measurements up to 9 T were conducted in Quantum Design PPMS-9 with conventional four-wire method. In the
in-plane resistivity and Hall measurements, the current path was in the \textit{bc}-plane, whereas magnetic field was parallel to the \textit{a}-axis except in the angular
dependent MR measurement. Seebeck coefficient was measured using steady state method and one-heater-two-thermometer setup with silver paint contact directly on the sample surface.
The heat and electrical current were transported within the \textit{bc}-plane of the crystal, with magnetic field along the \textit{a}-axis and perpendicular to the heat/electrical
current. The relative error in our measurement for both $\kappa$ and $S$ was below $5\%$ based on Ni standard measured under identical conditions. The de Haas-van Alphen (dHvA) oscillation experiments were performed at National High Magnetic Field Laboratory, Tallahassee. The crystals were mounted onto miniature Seiko piezoresistive cantilevers which
were installed on a rotating platform. The field direction can be changed continuously between parallel ($\theta=0^o$) and perpendicular ($\theta=90^o$) to the $c$-axis of the crystal.

\section{Results and discussions}

Sb$_2$Te$_2$Se has tetradymite structure consisting of three quintuple layers and can be represented as -Te-Sb-Se-Sb-Te-Te-Sb-Se-Sb-Te- (as shown in the inset of Fig. 1(b)).
Powder XRD pattern of our crystals which were fitted by RIETICA software~\cite{rietica} in Fig. 1(a) can be indexed in the R$\bar{3}$m space group. The crystals are
plate-like and the base plane is the $ab$-plane (Fig. 1(b)). SEM elementary analysis revealed that the composition of our crystals is Sb$_{2.02(7)}$Te$_{1.95(3)}$Se$_{1.05(7)}$.
The in-plane resistivity of the crystal is metallic above $\sim 20$ K (Fig. 2(a)) and then increases with decreasing temperature below 20 K (as shown in the inset of Fig. 2(a)).
The external magnetic field enhances the resistivity in the whole temperature range significantly, while it has barely any influence on the thermal conductivity (Fig. 2(b)).
The thermopower behavior of Sb$_2$Te$_2$Se is interesting (Fig. 2(c)). The Seebeck coefficient is positive above $\sim 100$ K. But it decreases with decreasing temperature and
becomes negative below 100 K. The magnetic field induces the decrease of the sign reversal temperature of Seebeck coefficient.

\begin{figure}[tbp]
\includegraphics[scale=0.4]{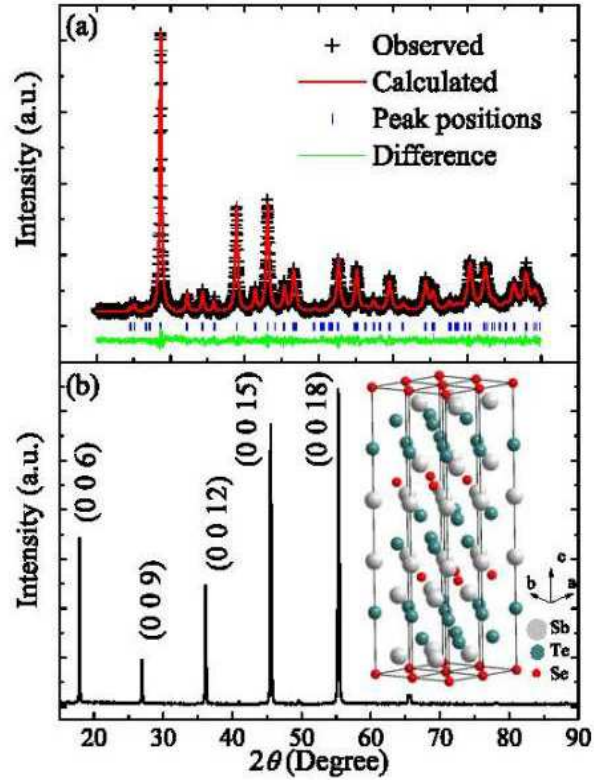}
\caption{(Color online) (a) Powder XRD patterns and structural refinement results. The data were shown by ($+$) , and the fit is given by the red solid line. The difference curve (the green solid line) is offset. (b) Single crystal XRD pattern shows that the basal plane of a crystal is the $ab$-plane. The inset of (b) shows the crystal structure of Sb$_2$Te$_2$Se.}
\end{figure}

\begin{figure}[tbp]
\includegraphics[scale=1] {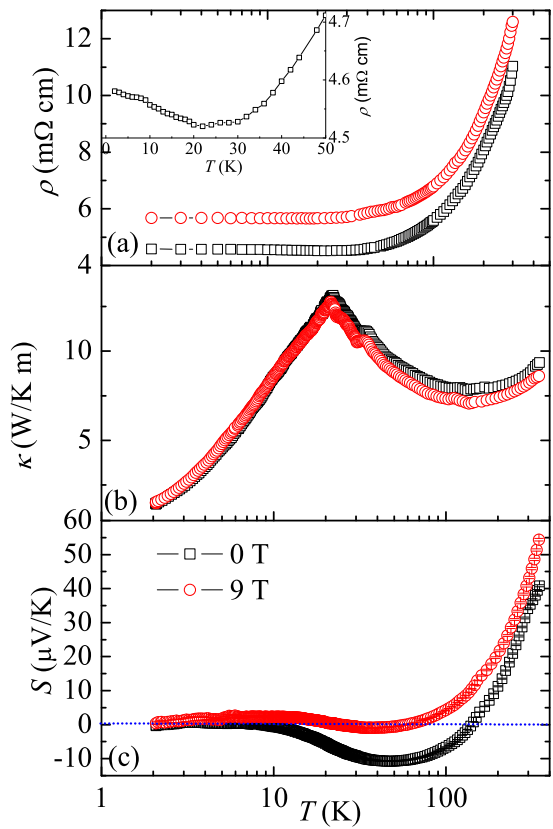}
\caption{(Color online) In-plane resistivity $\rho_{ab}(T)$ (a), thermal conductivity $\kappa(T)$ (b) and Seebeck coefficient $S(T)$ (c) of Sb$_2$Te$_2$Se single crystal as a function of temperature in 0 T and 9 T magnetic field respectively. The inset in (a) shows the magnified part of the semiconductor-metal transition around 25 K.}
\end{figure}

Below we will discuss the magnetic field effects on the resistivity and Seebeck coefficient. Fig. 3(a) shows the magnetic field dependence of the MR at several
temperature. At 2 K, the MR reaches $\sim 30\%$ in 9 T field (Fig. 3 (a)). In Fig. 3(b), the field derivative of MR, $d$MR$/dB$, initially decreases with increase in field indicating $B^{1/2}$ dependence of MR,
and then linearly increases with field in the low field region which indicates a $B^2$ dependent MR by linear fitting (lines in the low field region).
But above a characteristic field $B^*$, $d$MR$/dB$ saturates to a much reduced slope  This indicates that in the high fields the MR is dominated by a linear field dependence
plus a very small quadratic term (MR $=A_1B+O(B^2)$) as shown by lines in the high-field region, which extends to a very low crossover fields $B^*$ where the MR naturally reduces to a weak-field semiclassical
quadratic response. The high field linear MR persists even at 300 K. Similar room temperature linear
MR was observed in Bi$_2$Te$_3$ nanosheet.\cite{QT1}

\begin{figure}[tbp]
\includegraphics[scale=1] {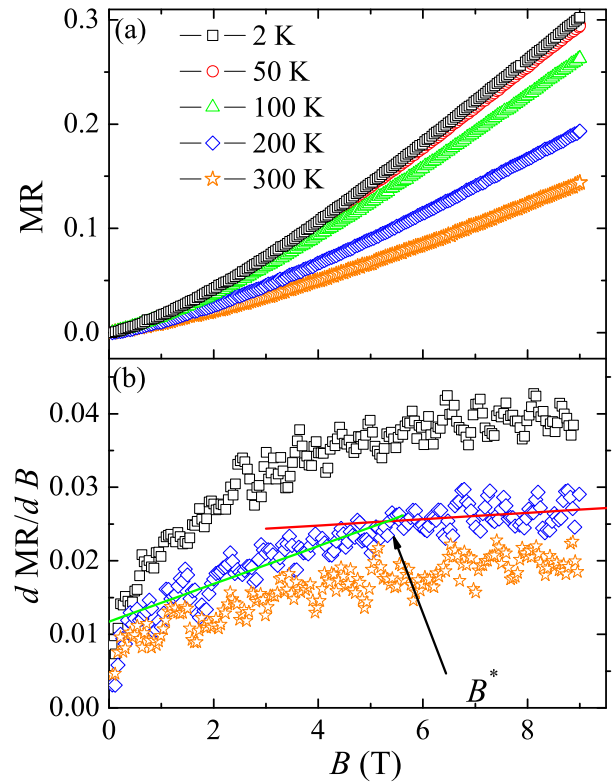}
\caption{(Color online) (a) The magnetic field ($B$) dependence of the in-plane magnetoresistance MR at different
temperatures. (b) The field derivative of in-plane MR, $d$MR$/dB$, as a function of field (B) at different temperature
respectively. The red lines in high field regions were fitting results using MR $=A_1B+O(B^2)$ and the lines in low field regions using MR $=A_2B^2$}
\end{figure}

\begin{figure}[tbp]
\includegraphics[scale=1] {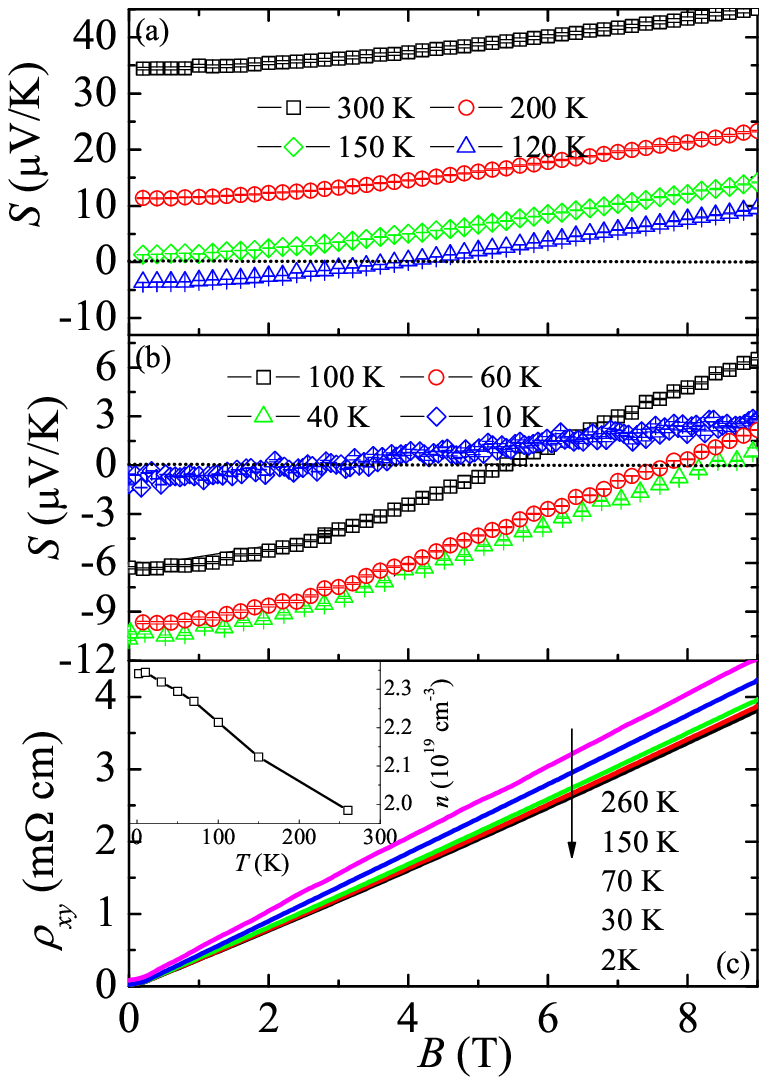}
\caption{(Color online)(a,b) The magnetic field ($B$) dependence of the Seebeck coefficient ($S$) at several different temperatures. The dot lines show the position of the zero Seebeck coefficient and indicate the sign change of the Seebeck coefficient. (c) Hall resistivity as a function of the magnetic field at different temperatures. The inset shows the temperature dependence of the carrier density deduced from Hall resistivity.}
\end{figure}

The linear magnetic field response was also found in the Seebeck coefficient (Fig. 4(a) and (b)). As shown in Fig. 2(c) and Fig. 4(a), the magnetic field enhances the Seebeck
coefficient above 150 K but the Seebeck coefficient is positive in whole field range. Below $\sim 120$ K, the Seebeck coefficient in zero field is negative in low fields where the absolute value of S decreases linearly with increase in magnetic field. Above specific crossover field there is a sign change of S from negative to positive where the absolute value of $S$ increases with increase in field. At 120 K, the crossover field is around 3 T. With decreasing
temperature, the crossover field initially increases to about 8 T at 40 K and then decreases to around 2 T at 10 K.

Fig. 4(c) shows the Hall resistivity and the apparent carrier density as function of temperature and field. The Hall resistivity exhibits linear field dependence,
the signature of single-band behavior. But contrary to the sign change of Seebeck coefficient with changing temperature and field, the Hall resisitivity for Sb$_2$Te$_2$Se
is always linear and positive in whole temperature and field range (Fig. 4(c)). The apparent carrier density $n_{app}=B/(e\rho_{xy})$ increases with decreasing temperature which is
consistent with the metallic behavior (the inset in Fig. 4(c)).

\begin{figure}[tbp]
\includegraphics [scale=0.4]{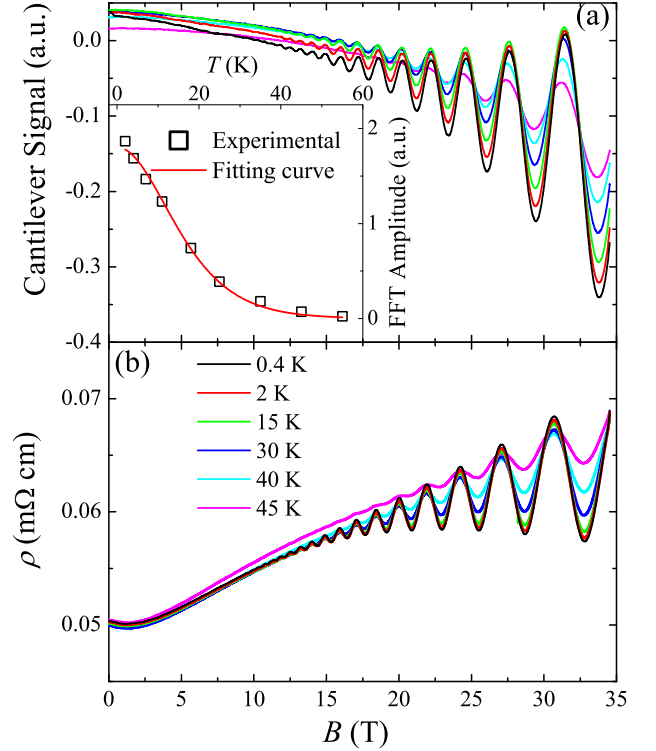}
\caption{(Color online) Cantilever (a) and magnetoresistant (b) oscillation as a function of field below 35 T at different temperatures. The inset of (a) shows the temperature dependence of the oscillation amplitude of the Fourier transform spectrum (FFT Amplitude) in cantilever oscillations. The red line is the fitting results giving cyclotron mass.}
\end{figure}


\begin{figure}[tbp]
\includegraphics [scale=1]{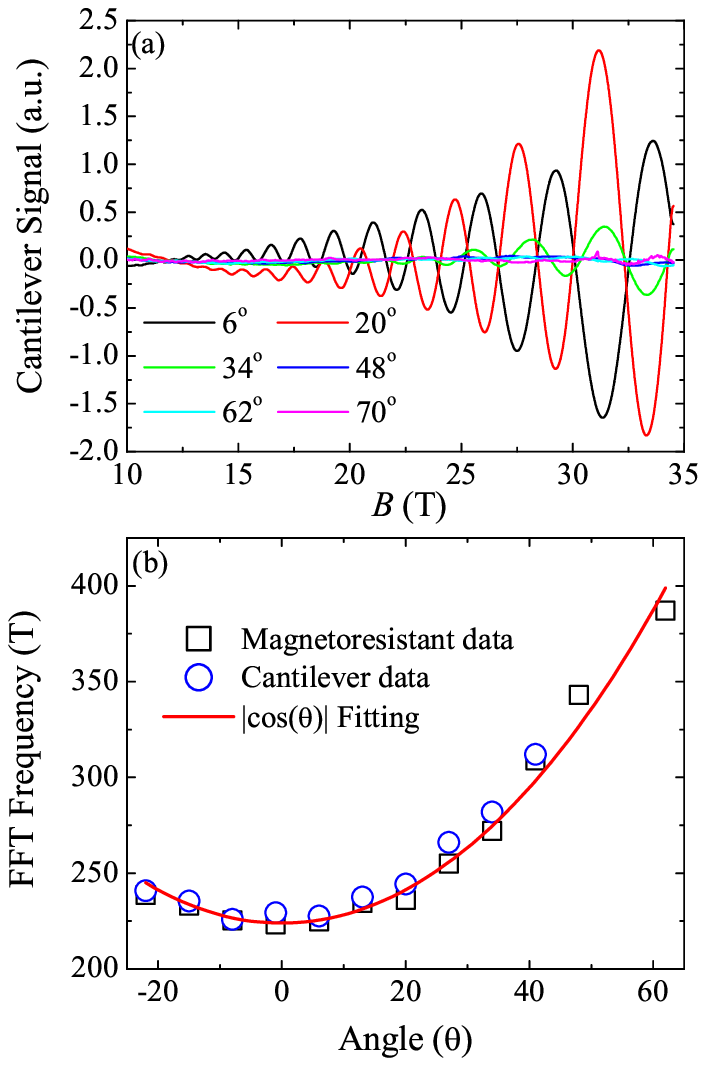}
\caption{(Color online) (a) Cantilever oscillation as a function of field  at different tilting angles $\theta$. (b) The angular dependence of the quantum oscillation frequency. Both of the data from magnetoresistance SdH oscillation (open squares) and cantilever oscillation (open circles) are presented. The red line is the fitting result using $1/|\cos(\theta)|$.}
\end{figure}

The tetradymite-like compounds such as Bi$_2$Te$_2$Se, Bi$_2$Te$_2$S, Bi$_2$Se$_2$S and Sb$_2$Te$_2$Se were predicted to host an isolated Dirac cone on their naturally
cleaved surface and attracted intensive attention.\cite{221-1,221-2,221-3,221-4} In order to clarify the possible topological surface state and the electronic structure,
we performed the quantum oscillation measurements up to 35 T field on Sb$_2$Te$_2$Se crystals. Both the magnetoresistance and the magnetic torque shows quantum oscillation.
The temperature-dependent and angular ($\theta$) dependent cantilever or magnetoresistance signals are shown in Fig. 5 and Fig. 6, respectively. The Fourier transform spectrum of the oscillation
at 0.4 K reveals a  periodic behavior in $1/B$ with a frequency $F\sim 225$ T when field is perpendicular to the $ab$-plane ($\theta$=0), as shown in Fig. 6(b).
The temperature dependence of the oscillation amplitude can be used to determine cyclotron effective mass through the Lifshitz-Kosevitch formula.\cite{oscillation}
Using the highest oscillation peak, the fitting gives a $m\approx 0.13(5) m_e$ where $m_e$ is the bare electron mass (the inset of Fig. 5(a)).
Fig. 6(a) shows the magnetic field direction dependence of the typical dHvA oscillations of Sb$_2$Te$_2$Se single crystal, and the detailed angular dependence
of the oscillation frequency are shown by squares in Fig. 6(b). For a surface state or two-dimensional Fermi surface, the cross section has $S_F(\theta)=S_0/|\cos(\theta)|$
angular dependence and the oscillation frequencies should be inversely proportional to $|cos(\theta)|$. In Fig. 6(b), the angular dependence of the oscillation frequency
can be fitted very well by $1/|\cos(\theta)|$ (the solid line). This indicates that the observed states in the quantum oscillation is quasi-2D Fermi surface.

However, the low-temperature resistivity of our material is $\sim 5$ m$\Omega$ cm and the density of carrier from the Hall resistivity is about $2\times10^{19}$ cm$^{-3}$, and both values are close to that observed in doped TIs such as Bi$_2$Se$_3$ and Bi$_2$Te$_3$. This means the bulk states should dominate the transport behavior in this material and the quasi-two-dimensional Fermi surface observed in quantum oscillation should not come from the Dirac surface states. In highly doped Bi$_2$Se$_3$ with a carrier density $\sim 4.7\times 10^{19}$ cm$^{-3}$, it was reported that the Fermi surface is large enough to touch the Brillouin zone boundary. So the quasi-2D magnetotransport features in this system arise from the bulk of the sample acting as many parallel 2D electron system to give a multilayered quantum Hall effect, instead of the Surface state.\cite{metal1} Our material has similar carrier density and then the quais-2D magnetotransport behavior observed in quantum oscillation (Fig. 6) should have similar origin. Besides that, the band structure calculation reveals that the bulk state in Se-doped Sb$_2$Te$_3$ most likely is massive Dirac state.\cite{221-1,221-2}, which possibly induce the linear magnetiresistance in high field as observed in other materials with Dirac states.\cite{QT1,QT2}

In a nonmagnetic metal, the diffusion of carriers and the phonon drag effect will contribute to the Seebeck coefficient. The diffusion mechanism of electron and holes will determine the sign of the Seebeck coefficient. The phonon drag mechanism often gives a peak structure at temperature $T\sim \Theta_D/5$ where $\Theta_D$ is the Debye temperature.\cite{TE1,TE2} In our crystal, the Debye temperature from the fitting the specific data is $\sim 185$ K, and the peak around $50$ K in both of the Seebeck coefficient in 0 T and 9 T field [Fig. 2(c)] should come from the phonon drag. But the sign change of $S$ below 10 K cannot originate from the phonon drag effect since magnetic field should have no influence on the phonon and the Seebeck coefficient below 10 K is linearly temperature-dependent which implies that the diffusion mechanism dominates the low temperature Seebeck coefficient. In a single band metal with diffusion mechanism and electron-type carriers, Seebeck coefficient is given by the Mott relationship,
\begin{eqnarray}
S(B)&=&\mathcal{A}\left(\frac{\sigma^2}{\sigma^2+\sigma_{xy}^2}\mathcal{D}+\frac{\sigma_{xy}^2}{\sigma^2+\sigma_{xy}^2}\mathcal{D}_H\right)\\
&=&\mathcal{A}\frac{(Ne\mu)^2\mathcal{D}+(Ne\mu^2B)^2\mathcal{D}_H}{(Ne\mu)^2+(Ne\mu^2B)^2};
\end{eqnarray}
where $\mathcal{A}=\frac{\pi^2k_B^2T}{3e}$, $\mathcal{D}=\frac{\partial\ln\sigma}{\partial \zeta}$ and $\mathcal{D}_H=\frac{\partial\ln\sigma_{xy}}{\partial \zeta}$ ($\zeta$ is the chemical potential).\cite{TE1,TE2,TE3} From it the dependence of Seebeck coefficient on field $B$ appears only in the conductivity matrix element $\sigma_{ij}(B)$  ($\sigma=\sigma_{xx}$). The electron contribution to Seebeck coefficient $S_e$ is usually negative while the hole contribution $S_h$ is always positive.\cite{TE1,TE2} For a two-band metal comprising electron and hole bands, $S$ is expressed as
\begin{eqnarray}
S=\frac{\sigma_h|S_h|-\sigma_e|S_e|}{\sigma_h+\sigma_e},
\end{eqnarray}
where $\sigma_{e(h)}$  and $S_{e(h)}$ are the contributions of electrons (holes) to the electric conductivity and Seebeck coefficient, respectively.\cite{TE1} So the different sign of
Seebeck coefficient and Hall resistivity implies the multiband effect in Sb$_2$Te$_2$Se, while the linear positive Hall resistivity implies that the single hole band dominates the Hall transport.

Similar sign change in the Seebeck coefficient by decreasing temperature was also observed in Se-doped Bi$_2$Se$_3$ \cite{BiTeSe} and one possible reason for this and the different sign between the Hall and Seebeck coefficient is the negative phonon drag peak superposed by the positive diffusive thermoelectric response.\cite{TE1,TE2} This could explain the positive-negative transition of Seebeck coefficient by decreasing the temperature in our material. However, it could not induce the sign change by increasing magnetic field since the magnetic field should have no influence on the phonon drag, and this means the phenomena should come from the carriers. Similar phenomena where Hall resistivity and Seebeck coefficient has different sign were also observed in some noble metals such as Cu and Ag and was attributed to the neck structure of the Fermi surface.\cite{Cu} The Fermi surface in Ag and Cu is far from spherical, but Fermi surface just touches the Brillouin boundary and has a set of "necks" at the Brillouin boundary. \cite{FS} Our material has high carrier density and the quasi-2D magnetotransport behavior in quantum oscillation suggests that the Fermi surface (hole) in doped Sb$_2$Te$_2$Se is large enough to touch the Brillouin zone boundary and should give similar neck structure as Cu/Ag. Usually these necks are almost perpendicular to the Brillouin boundary and the curvatures near the necks are oppositive to the residual Fermi surface, as shown in Cu/Ag case. The neck structure gives a heavy electron-like (negative) contribution to the Seebeck coefficient. The density of this electron-like states associated with the necks, is extremely large due to the rapidly varying surface with energy. But its contribution to the Hall coefficient is very small because it has the very large effective mass and the residual Fermi surface is massive Dirac state with very large mobility. Consequently the different sign between the Hall and Seebeck coefficient appears.

The sign reversion of Seebeck coefficient in magnetic field should be related to the different response of two kind of contribution from different parts of the Fermi surface. Similar effects were also reported in several other materials such as high-$T_c$ cuprates, and was considered as a signal of the Fermi surface reconstruction.\cite{cuprate1,cuprate2} For example, the high-field $R_H$
and $S/T$ are found to drop with decreasing temperature and become negative in HgBa$_2$CuO$_{4+\delta}$, which is attributed to the charge-order wave and related Fermi surface reconstruction.\cite{cuprate2} In similar materials Bi$_2$Se$_3$ with high bulk carrier density, a strong variation of the chemical potential and Fermi energy above the quantum limit is observed and is believed to induce the observed increase in the Nernst response.\cite{behina} The change of the chemical potential could also happen in Sb$_2$Te$_2$Se and it could suppress the contribution of the states at the Necks of the Fermi surface to the Seebeck coefficient by suppressing the number of that states. Besides that,with increasing magnetic field, the Dirac holes will be very easy to form Landau splitting and then occupy the zeroth LLs (achieve quantum limit) gradually because of the high mobility and small effective mass, while the moderate field hardly has any influence on the states at the necks due to the very large mass. So under magnetic field the Dirac holes will dominate the thermal transport behavior in the quantum limit since the Fermi level
locates between the zeroth and first LLs and the DOS at the Fermi level is suppressed. The positive Hall resistivity and Seebeck coefficient confirm the dominant hole-like carriers up to some crossover field above which there is a sign change in $S(B)$ due to enhanced contribution of bulk hole-like carriers.

\section{Conclusion}
In summary, we report the magnetoresistance, magnetothermopower and quantum oscillation study of Sb$_2$Te$_2$Se single crystal.  The in-plane transverse magnetoresistance
exhibits a crossover at a critical field $B^*$ from semiclassical weak-field $B^2$ dependence to the high-field unsaturated linear magnetoresistance which persists up to the
room temperature. The low-temperature Seebeck coefficient is negative in zero field contrary
to the positive Hall resistivity, indicating the multiband effect. The magnetic field induced the sign reversion of the Seebeck coefficient between 2 K and 150 K,
. The quantum oscillation of crystals
reveals the quasi-2D Fermi surface. These effects are possibly attributed to the large Fermi surface which touches Brillouin zone boundary to become quasi-2D and the variation in the chemical potential induced by the magnetic field.

\begin{acknowledgments}
We than John Warren for help with SEM measurements. Work at Brookhaven is supported by the U.S. DOE under contract No. DE-AC02-98CH10886. The high magnetic field studies in NHMFL were supported by NSF DMR-0654118, the State of Florida and DOE NNSA DE-FG52-10NA29659.
\end{acknowledgments}


\begin{thebibliography}{99}
\bibitem{thermoelectric1}
D. M. Rowa, \textit{Thermoelectrics Handbook: Macro to Nano} (CRC Press, Boca Rotan, 2006).
\bibitem{thermoelectric2}
G. S. Nolas, J. Sharp, and H. J. Goldsmid, \textit{Thermoelectrics: Basic  Principles and New Materials Developments} (Springer, New York, 2001).
\bibitem{TI1}
X. L. Qi and S. C. Zhang, Rev. Mod. Phys. {\bf 83}, 1057 (2011).
\bibitem{TI2}
M. Z. Hasan and C. L. Kane, Rev. Mod. Phys. {\bf 82}, 3045 (2010).
\bibitem{TI3}
L. Fu, C. L. Kane, and E. J. Mele, Phys. Rev. Lett. {\bf 98}, 106803 (2007).
\bibitem{TI4}
D. Qu, Y. S. Hor, J. Xiong, R. J. Cava and N. P. Ong, Science {\bf 329}, 821 (2010).
\bibitem{TI5}
J. G. Analytis, R. D. McDonald, S. C. Riggs, J.-H. Chu, G. S. Boebinger, and I. R. Fisher, Nature Phys. {\bf 6}, 960 (2010).
\bibitem{TI6}
D. Hsieh, D. Qian, L. Wray, Y. Xia, Y. S. Hor, R. J. Cava, and M. Z. Hasan, Nature {\bf 452}, 970 (2008).
\bibitem{TI7}
Y. L. Chen, J. G. Analystis, J. H. Chu, Z. K. Liu, S. K. Mo, X. L. Qi, H. J. Zhang, D. H. Lu, X. Dai, Z. Fang, S. C. Zhang, I. R. Fisher, Z. Hussain, and Z. X. Shen, Science {\bf 325}, 178 (2009).
\bibitem{TI8}
D. Xiao, Y. Yao, W. Feng, W. Zhu, X. Q. Chen, G. M. Stocks, and Z. Y. Zhang, Phys. Rev. Lett. {\bf 105}, 096404 (2010).
\bibitem{TI9}
B. A. Bernevig, T. L. Hughes, and S. C. Zhang, Science {\bf 314}, 1757 (2006).
\bibitem{TI10}
Y. Xia, et al. Nature Phys. {\bf 5}, 438 (2009).
\bibitem{221-1}
H. Lin, T. Das, L. A. Wray, S.-Y. Xu, M. Z. Hasan, and A. Bansil, New. J. Phys. {\bf 13}, 095005 (2011).
\bibitem{221-2}
W. Zhang, et al. New. J. Phys. {\bf 12}, 065013 (2010).
\bibitem{221-3}
Z. Ren, A. A. Taskin, S. Sasaki, K. Segawa, and Y. Ando, Phys. Rev. B {\bf 84}, 165311 (2011).
\bibitem{221-4}
S.-Y. Xu, et al. arXiv:1007.5111
\bibitem{QT1}
X. L. Wang, Y. Du, S. X. Dou and C. Zhang, Phys. Rev. Lett. {\bf 108}, 266806 (2012).
\bibitem{QT2}
D. Qu, Y. S. Hor, and R. J. Cava, Phys. Rev. Lett. {\bf 109}, 246602 (2012).
\bibitem{metal1}
H. Cao, J. Tian, I. Miotkowski, T. Shen, J. Hu, S. Qiao, and Y. P. Chen, Phys. Rev. Lett. {\bf 108}, 216803 (2012).
\bibitem{metal2}
Z. J. Xue, C. B. Zhu, S. X. Dou, and X. L. Wang, Phys. Rev. B {\bf 86}, 195120 (2012).
\bibitem{behnia}
B. Fauqu$\acute{e}$, N. P. Butch, P. Syers, J. Paglione, S. Wiedmann, A. Collaudin, B. Grena, U. Zeitler, and K. Behnia, Phys. Rev. B {\bf 87}, 035133 (2013).
\bibitem{InSb}
J. P. Heremans, C. M. Trush, and D. T. Morelli, Phys. Rev. Lett. \textbf{86}, 2098 (2001).

\bibitem{gmr}
J. Shi, K. Pettit, E. Kita, S. S. P. Parkin, R. Nakatani, and M. B. Salamon, Phys. Rev. B {\bf 54}, 15273 (1996).

\bibitem{cmr}
M. Jaime, M. B. Salamon, K. Pettit, M. Rubinstein, R. E. Treece, J. S. Horwitz, and D. B. Chrisey, Appl. Phys. Lett. {\bf 68}, 1576 (1996).

\bibitem{rietica}
B. Hunter, "RIETICA - A Visual RIETVELD Progarm," International Union of Crystallography Commission on Powder Diffractin Newsletter No. 20 (Summer), 1998 (http://www.rietica.org).

\bibitem{oscillation} D. Shoeneberg, \textit{Magnetic oscillation in metals} (Cambridge University Press, Cambridge, 1984).

\bibitem{LL1} Y. Zhang, Z. Jiang, Y.-W. Tan, H. L. Stormer, and P. Kim,
Nature \textbf{438}, 201 (2005).

\bibitem{LL2} D. Miller, K. Kubista, G. Rutter, M. Ruan, W. de Heer, P.
First, and J. Stroscio, Science \textbf{324}, 924 (2009).

\bibitem{agte1}
R. Xu, A. Husmann, T. F. Rosenbaum, M.-L. Saboung, J. E. Enderby, and P. B. Littlewood, Nature {\bf 390}, 57 (1997).

\bibitem{agte2}
W. Zhang, R. Yu, W. Feng, Y. Yao, H. Wang, X. Dai, and
Z. Fang, Phys. Rev. Lett. \textbf{106}, 156808 (2011).

\bibitem{agte3} W. Zhang, R. Yu, W. Feng, Y. Yao, H. Wang, X. Dai, and
Z. Fang, Phys. Rev. Lett. \textbf{106}, 156808 (2011).

\bibitem{122-1} K. K. Huynh, Y. Tanabe, and K. Tanigaki, Phys. Rev. Lett.
\textbf{106}, 217004 (2011).

\bibitem{122-2} H.-H. Kuo, J.-H. Chu, S. C. Riggs, L. Yu, P. L. McMahon, K. D.
Greve, Y. Yamamoto, J. G. Analytis, and I. R. Fisher, Phys. Rev. B {\bf 84}, 054540 (2011).

\bibitem{srmnbi21} J. Park, G. Lee, F. Wolff-Fabris, Y. Y. Koh, M. J. Eom, Y.
K. Kim, M. A. Farhan, Y. J. Jo, C. Kim, J. H. Shim, and J. S. Kim, Phys. Rev. Lett. {\bf 107}, 126402 (2011).

\bibitem{srmnbi22}
K. Wang, D. Graf, H. Lei, S. W. Tozer, and C. Petrovic, Phys. Rev. B {\bf 84}, 220401(R) (2011).

\bibitem{srmnbi23}
K. Wang, D. Graf, L. Wang, H. Lei, S. W. Tozer, and C. Petrovic, Phys. Rev. B {\bf 85}, 041101(R) (2012).

\bibitem{qmr1}
A. A. Abrikosov, Phys. Rev. B \textbf{58}, 2788 (1998).

\bibitem{qmr2}
A. A. Abrikosov, Europhys. Lett. \textbf{49}, 789 (2000).

\bibitem{mr3}
A. B. Pippard, \textit{Magnetoresistance in Metals} (Cambridge University, Cambridge, 1989).

\bibitem{TE1} R. D. Barnard, \textit{Thermoelectricity in Metas and Alloys} (Taylor \& Francis, London, 1972).

\bibitem{TE2} J. M. Ziman, \textit{Electrons and Phonons} page 500 (Oxford Clarendon Press, Oxford, 1960).

\bibitem{TE3} T. Liang, Q. Gibson, J. Xiong, M. Hirschberger, S. P. Koduvayur, R. J. Cava and N. P. Ong, Nat. Commun. {\bf 4}, 2696 (2013).

\bibitem{BiTeSe} A. Akrap, A. Ubaldini, E. Giannini, and L. Forr\'{o}, arXiv:1210.3901.

\bibitem{Cu} S. Fujita, H.-C. Ho, and Y. Okamura, Int. J. Mod. Phys. B \textbf{14}, 2231 (2000).

\bibitem{FS} D. Sch\"{o}nberg, Phil. Trans. R. Soc. Lond. \textbf{255}, 85 (1962).

\bibitem{cuprate1} F. Laliberte, J. Chang, N. Doiron-Leyraud, E. Hassinger, R. Daou, M. Rondeau, B. J. Ramshaw, R. Liang, D. A. Bonn, W. N. Hardy, S. Pyon, T. Takayama, H. Takagi, I. Sheikin, L. Malone, C. Proust, K. Behnia, and L. Taillefer, Nat. Commun. {\bf 2}, 432 (2011)

\bibitem{cuprate2} N. Doiron-Leyraud, S. Lepault, O. Cyr-Choiniere, B. Vignolle, G. Grissonnanche, F. Laliberte, J. Chang, N. Barisic, M. K. Chan, L. Ji, X. Zhao, Y. Li, M. Greven, C. Proust, and Louis Taillefe, Phys. Rev. X {\bf 3}, 021019 (2013).




\end{thebibliography}

\end{document}